\documentstyle[aps,epsfig,prl,multicol]{revtex}
\begin{document}
\draft

\title{Ballistic Localization in Quasi-1D Waveguides with Rough Surfaces}

\author {F.M. Izrailev, J.A. M\'endez-Berm\'udez, and G.A. Luna-Acosta}
\address{Instituto de F\'{\i}sica, Universidad Aut\'onoma de Puebla,
Apdo. Postal J - 48, Puebla 72570,  Mexico.}

\date{\today}
\maketitle
\begin{abstract}

\end{abstract}

\begin{abstract}
Structure of eigenstates in a periodic quasi-1D waveguide
with a rough surface is studied both analytically and numerically.
We have found a large number of "regular" eigenstates
for any high energy. They result in a very
slow convergence to the classical limit in which the eigenstates are
expected to be completely ergodic. As a consequence, localization properties
of eigenstates originated from unperturbed transverse channels with low
indexes, are strongly localized (delocalized) in the momentum (coordinate)
representation. These eigenstates were found to have
a quite unexpeted form that manifests a kind of "repulsion" from the rough
surface. Our results indicate that standard statistical approaches for ballistic
localization in such waveguides seem to be unappropriate.
\end{abstract}


\begin{multicols}{2}


In the past decade much attention has been paid to statistical properties of
eigenstates of closed disordered systems. As a result, to date practically 
everything is known for quasi-1D systems with "bulk disorder".
The success 
is mainly related to the developments of the non-linear $\sigma$ model
(see, e.g. \cite{sigma} and references therein). One of the most important
results is that the statistical properties of such systems are essentially 
determined by one characteristic length only, known as the
localization length of eigenstates.
This fact is entirely due to a strong
mixing between transverse channels resulting from {\it bulk scattering}
and leading to a diffusive character of transport.

A much more difficult situation was found to occur for the models with
{\it surface scattering} when the disorder
is due to a surface roughness. In quasi-1D geometry
such models are closely related to optical/microwave waveguides and have many
physical applications in different fields \cite{ping}. The main 
problem in the rigorous 
treatment of this kind of systems is in the ballistic character of the scattering
which has weaker statistical properties in comparison with diffusive scattering.
Progress in this direction is related to recent developments of the "ballistic 
sigma-model", however, the problem is still far from being solved (see discussion 
and references in \cite{GM02}).

As was recently shown in a number of numerical studies \cite{recent,SFYM99},
the transport in quasi-1D waveguides with rough surfaces is highly 
non-isotropic in channel space. Specifically, 
the transport through such waveguides strongly depends on the
incident angle of incoming wave. In particular, the transmission 
coefficient smoothly decreases with an increase of the angle, since
characterictic lengths for backscattering are different for different
channels \cite{SFYM99}.

To understand generic features of surface disordered quasi-1D systems, 
in this Letter we perform a detailed study of the structure 
of eigenstates of a 2D quantum billiard (or
waveguide). We consider billiards
which are periodic in the longitudinal coordinate $x$, 
and with Dirichlet
boundary conditions on the low, $y=0$, and upper,
$y=d+a\xi(x)$, surfaces 
with $\xi(x+b)=\xi(x)$ and $\left<
\xi(x)\right>=0$.
Here the angular brackets stand for the average over one period $b$
(or, in the case of a random profile,
over different realizations of $\xi(x)$).

Our main interest is in the study of the structure of eigenstates
of this billiard in dependence on their energy and model
parameters. For this we use the technique that transforms the
Hamiltonian $\hat H =\frac{1}{2m_e}(\hat P_x^2 + \hat P_y^2)$ for a
free particle inside the billiard with the above boundaries, to
a new Hamiltonian which incorporates surface scattering effects
into effective interaction potential. This can be achieved by the
transformation to new canonical coordinates, $u = x;\,\,\, v =
\frac{y}{1+\epsilon \xi(x)}$ with $\epsilon = a/d$. As a result,
the boundary conditions for new wave function are trivial:
$\Phi(u,v)=0$ at $v=0$ and $v=d$ (see details in \cite{qrch})

In the new variables the Hamiltonian gets \cite{qrch},
\begin{equation}
\hat H= -\frac{\hbar^2}{2m_e}\left(\frac{\partial^2}{\partial u^2} +
h_1
\frac{\partial^2}{\partial v^2} + h_2\frac{\partial^2}{\partial u
\partial v} + h_3
\frac{\partial}{\partial v}\right),
\label{Ham1}
\end{equation}
where
\begin{equation}
\begin{array}{ll}
h_1 = \frac{1 + \epsilon^2 v^2 \xi_u^2}{(1+\epsilon \xi)^2},
\,\,\,\, h_2 =
\frac{-2 \epsilon v \xi_u}{1+\epsilon \xi},\,\,\,
h_3 = \frac{-\epsilon v \xi_{uu}}{1 +\epsilon \xi}+
\frac{2v \epsilon^2\xi_u^2}{(1+\epsilon \xi)^2},
\label{hh}
\end{array}
\end{equation}
and $\xi_u = \partial \xi /\partial u, \, \xi_{uu} = \partial^2 \xi
/\partial u^2$.

One can write the Hamiltonian in the following form,
\begin{equation}
\hat H = \hat H^0 + \hat V(u,v, \hat P_u, \hat P_v)\,;\,\,\,\,\,
\hat H^0=\frac{1}{2m_e}(\hat P_u^2 + \hat P_v^2)
\label{H00}
\end{equation}
where $\hat P_u$ and $\hat P_v$ are the new canonical momenta.
In this way, the "unperturbed" Hamiltonian $\hat H^0$ descibes free motion
of two "particles" inside a billiard 
with flat boundaries, $y=0,\,\,\, y=d$, and
$\hat V$ stands for an effective ``interaction" between the
``particles". Such a representation turns out to be very
convenient for
the study of chaotic properties of our model, since one can use
the tools and concepts developed in the theory of interacting
particles (see \cite{I01} and references therein).

This model has been thoroughly studied in 
Refs.\cite{qrch,Pla2000,crch,rmf,PE2002,others} for the
specific case $\xi(x)=\cos(2\pi x/b)$. The main
interest was in the properties of energy spectrum
\cite{qrch,rmf}, and in the quantum-classical correspondence for
the shape of eigenfunctions (SEF) and local density of states
(LDOS) \cite{Pla2000,PE2002}. In particular, it was shown that for
highly
excited states the global properties of the SEF and LDOS in the
quantum model (\ref{H00}) are similar to those described by the
equations of motion for a classical particle moving inside the
billiard. On the other hand, quite strong quantum effects have
been revealed for individual eigenstates in a deep
semiclassical region \cite{PE2002}.

Below we address the case of a rough surface 
$\xi(x) = \sum_{k=1}^{N_T} A_k \cos(2\pi kx/b)$, focusing
on the properties of eigenstates. 
The surface is modeled by a large sum of harmonics with
randomly distributed amplitudes $A_k$. 
With an increase of $N_T$ the degree of complexity of $\xi(x)$
increases and for a large $N_T \gtrsim 100$ the surface can be
treated as the random one.

Since the Hamiltonian (\ref{H00}) is periodic in $u=x$, the
eigenstates are Bloch states and the solution of the Schr\"odinger
equation can be written in the form $\psi_E(u,v)=\exp
(i\chi)\,\psi_\chi(u,v)$, with $\psi_\chi(u+2\pi /b,v)=\psi_\chi(u,v)$.
Here the Bloch wave vector $\chi(E)$ is in the first Brillouin band,
($-\frac{\pi}{b}\leq \chi \leq \frac{\pi}{b})$. By
expanding $\psi_\chi(u,v)$ in the basis of $\hat H_0$, the
$\alpha$-th eigenstate of energy $E^{\alpha}(\chi)$ can be written
as
\begin{equation}
\psi^{\alpha}(u,v;\chi)= \sum_{m=1}^{\infty}
\sum_{n=-\infty}^{\infty}C_{mn}^{\alpha}(\chi) \phi_{mn}^\chi(u,v),
\label{exp}
\end{equation}
where $\phi_{mn}^\chi(u,v)\,= \pi^{-1/2} g^{-1/4} \sin(m \pi
v/d)\,\exp\{i(\chi+2\pi n/b)u\}$. The factor $g=[1+\epsilon \xi(u)]^2$
arises from the orthonormality condition in the curvilinear
coordinates $(u,v)$ (see details in \cite{qrch}).

In the "unperturbed" basis defined by $\hat H^0$, the matrix
elements of the "interaction" $\hat V(u,v, \hat P_u, \hat P_v)$ can be
written explicitly for any profile $\xi(x)$ \cite{PE2002}. This fact
is very useful in the study of the dependence of the properties of
eigenstates on the form of profile.

The eigenvalues of $\hat H^0$ are given by the expression,
\begin{equation}
E^{(0)}_{n,m}(\chi) =
\frac{\hbar^2}{2m_e}\left[
\left( \chi+\frac{2\pi}{b}n \right)^2+\left(\frac{m\pi}{d}\right)^2\right].
\label{E00}
\end{equation}
In numerical simulations we have to make a cutoff for the values
of $m$ and $n$ in the expansion (\ref{exp}). Our main results refer to the ranges, $1 \leq m
\leq M_{max}$ and $|n| \leq N_{max}$ with $N_{max}=32
\,,M_{max}=62$, for which the total size of the Hamiltonian matrix is
$L=(2N_{max}+1)M_{max}=4030$. Since the statistical properties of
eigenstates do not depend on a specific value of the Bloch index
$\chi$ inside the band \cite{rmf}, we fix it to $\chi=0.1\pi/b$.

One natural representation of the Hamiltonian matrix
$H_{l,l'}(\chi)= <l\mid\hat H\mid l'>_\chi$ is the so-called
"channel representation" for which one fixes the values of $n$
starting from the lowest one, $n=-N_{max}$, with the run over $m$
for each $n$. In this way the matrix has a block structure
that manifests peculiarities of the interaction
between different channels specified by the index $m$. For our purpose, however, 
to analyze the properties of the eigenfunctions it is more convenient to
use the ``energy
representation" according to which the unperturbed basis is
ordered in increasing energy, $E^0_{l+1}\geq E^0_l$ \cite{PE2002}.

In what follows we mainly discuss periodic billiards with a weak
roughness, $\epsilon= a/d \ll 1$. However, all
matrix elements of the Hamiltonian are computed according to exact
analytical expressions. This is important because the
contribution of the "gradient" terms (that depend on $\xi_u, \, \xi_{uu}$)
is strong for $N\ll 1$ and should be
treated non-perturbatively. 

In order to characterize quantitatively the structure of eigenfunctions 
we compute the {\it entropy localization length}
$l_H$, given by the expression,
\begin{equation}
l_H =  \exp { \{ -\left( \cal H - \cal H_{GOE} \right)
\}} \approx 2.08 \exp \{-\cal H \} .
\label{shannon}
\end{equation}
Here ${\cal H} =\sum^L_{l=1} w_l^{\alpha} \,\ln w_l^{\alpha}$
stands for the Shannon entropy of an eigenstate in a
given basis, and $\cal H_{GOE}$ is the entropy of a completely
chaotic state characterized by gaussian fluctuations of the
components $C_l^\alpha$ with the
variance $<w_l^\alpha>=|C_l^{\alpha}|^2=1/L$ \cite{I90}. 
Note, that the value of $l_H$ is 
proportional to the localization length $l_{ipr}$, defined via the
{\it inverse participation ratio}, $l_{ipr} = 3 {\cal P}^{-1}$
with ${\cal P}= \sum^L_{l=1} (w_l^\alpha)^2$ \cite{I90}. Both quantities give
an estimate of the effective number of components in exact
(perturbed) eigenstates.

\begin{figure}
\hspace{-0.0cm}
\epsfxsize 8cm
\epsfbox{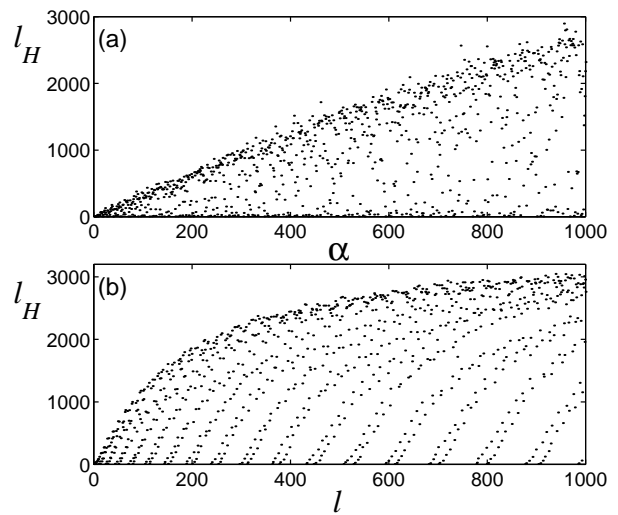}
\vspace{0.3cm}
\narrowtext
\caption{Localization measure $l_H$ for a rough surface with
$N_T=100$ for $\epsilon=0.06$ and $d=b$; (a) for exact
eigenstates $|~\alpha>$, (b) for individual LDOS $|~l>$.}
\end{figure}

In Fig. 1(a) the value of $l_H$ is plotted versus the index $\alpha$
for exact eigenstates ordered in energy $E^{\alpha}$. This {\it typical}
dependence of $l_H$ on $\alpha$ is quite instructive. As one can
anticipate, the number of principal components in the eigenstates increases,
in average, with energy. On the other hand, for any large energy
$E^{\alpha}$ there
are many eigenstates that have small values of $l_H$. For understanding 
the origin of these strongly {\it localized}
eigenstates (in the unperturbed basis $|~l>$), it is convenient to
consider the so-called {\it individual} LDOS \cite{Pla2000}. This quantity
corresponds to the representation of an unperturbed state $|~l>$
in the basis of exact states $|\alpha>$. Using the definition
(\ref{shannon}) with ${\cal H} =\sum^L_{\alpha =1} w_l^{\alpha}
\,\ln w_l^{\alpha}$ where the sum now runs over $\alpha$ for a specific
value of $l$, one can characterize
how many exact states contains specific unperturbed state $|~l>$.

The data of Fig. 1(b) show that there is a large number of
unperturbed states that seem to be close to the exact ones (with
$l_H \approx 1 $). The important point is that these states appear
in a {\it regular} way as a function of $l$. The analysis shows
that exact eigenstates $|~\alpha>$ with
smallest values of $l_H$ are originated from the unperturbed
states with $m=1$, see
Eq.(\ref{E00}). The data of Fig. 1(a) manifest that
localized eigenstates $|~\alpha> $ with small values of $l_H$ can be
classified by groups that are characterized by the values of
$m=1,2,3 \dots$ (for small $m$) of those unperturbed states $|~l>$ to
which they are "close".

\begin{figure}
\hspace{-0.0cm}
\epsfxsize 8cm
\epsfbox{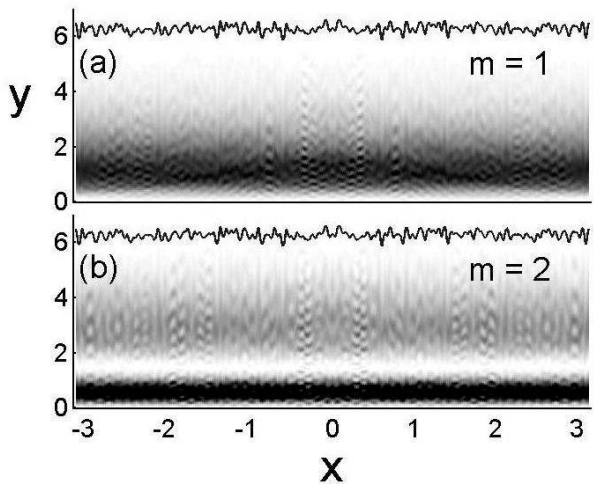}
\vspace{0.3cm}
\narrowtext
\caption{Examples of exact eigenstates that are strongly localized
in energy representation; (a) $\alpha=968$ ($m=1$), (b)
$\alpha=976$ ($m=2$). The probability
$|\psi^\alpha(x,y)|^2$ is plotted for the parameters of Fig. 1 (dark regions
correspond to high probablity). Broken
horizontal curves show upper surface profiles.}
\end{figure}

Two examples of such eigenstates are given in Fig. 2 in the coordinate
representation $(x,y)$. It is quite unexpected that these eigenstates are
very different from the unperturbed ones even though they are quite
close in energy representation. Fig. 2(a) shows that the rough
boundary ``pushes'' the probability $|\psi^\alpha(x,y)|^2$ away from it,
differing from the unperturbed mode with $m=1$ whose maximum is
at the center, $y=d/2$. Similar repulsion occurs for $m=2$,
see Fig. 2(b). Below we give the explanation of this
phenomenon by making use of the Hamiltonian in
variables $(u,v)$.

We start with the fact that eigenstates of the type
shown in Fig. 2(a) are exponentially localized in the $m-$space ({\it
channel space}) independently on $n$. Therefore, one can write,
$C_l^{\alpha}(k) = C_{mn}^{\alpha}(k) \sim \exp
\{-\beta(m-1)\}$ where $\beta$ is some constant (of the order of unity)
determined by numerical data, and surprisingly independent of energy.
Therefore, these localized eigenstates in the $(u,v)$ variables have the
following form,
\begin{equation}
\psi_{loc}^{\alpha} = \frac{C}{g^{1/4}} e^{
i(\chi+2\pi n_\alpha/b)u}
\sum_{m=1}^{\infty}  e^{ -\beta (m-1)} \sin \left(\frac{m \pi
v}{d}\right).
\label{loc}
\end{equation}

\noindent Here $C$ is the normalization constant determined by the
orthonormality condition in curvilinear coordinates,
$\int_0^{2\pi}
\int_0^1 du dv \sqrt{g} \psi_{loc}^{\alpha \ *}
\psi_{loc}^\alpha = 1$ (see \cite{qrch} for details).
As a result, one obtains,
\begin{equation}
\mid \psi_{loc}^{\alpha}(y) \mid ^2 =
\frac{e^{2\beta}-1}{4\pi}  \frac{\sin^2(\pi y/d)}
{\cosh \beta-\cos(\pi y/d)}
\label{square}
\end{equation}

\begin{figure}
\hspace{-0.0cm}
\epsfxsize 8cm
\epsfbox{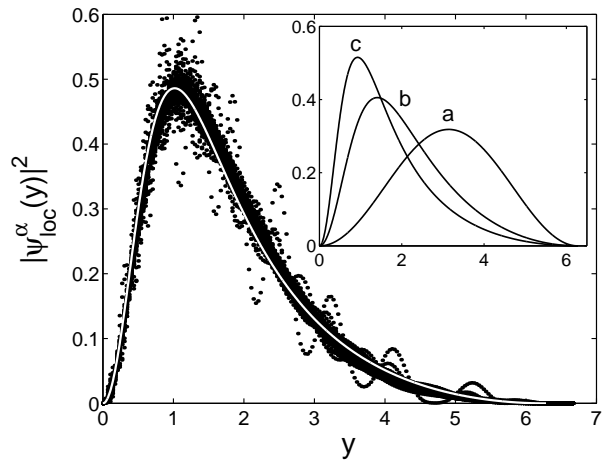}
\vspace{0.3cm}
\narrowtext
\caption{Projection of the eigenfuction profile of Fig. 2(a) onto the
$y-$coordinate (dots), together with the analytical expression
(\ref{square}) (white curve). The value $\beta=0.53$ was
numerically found by fitting $C_{mn}^{\alpha}(k)$ to the
exponential dependence $\exp\{-\beta(m-1)\}$. In the inset,
profiles (\ref{square}) for strongly localized eigenstates with $m=1$ in
dependence on the number $N_T$ of harmonics are shown for (a) $N_T=1$,
$\beta \rightarrow \infty$, (b) $N_T=50$, $\beta=0.77$, and
(c) $N_T \rightarrow \infty$, $\beta \rightarrow 0.48$.}
\end{figure}

The comparison of this expression with the numerical data is shown
in Fig. 3. One can see that in spite of a relatively weak coupling of low channels
($m \sim 1$) to all others, the scattering from a rough surface
strongly modifies the unperturbed states in $y-$direction. One can
speak about a kind of ``repulsion" of such eigenstates from the
rough surface. In order to see how this repulsion depends
on the degree of roughness, we have studied the form of
the states with $m=1$ in dependence on the number $N_T$ of
harmonics in the surface profile. The results
shown in the inset, reveal that with an increase of $N_T$ the
position of the maximum of the probability shifts away from the
rough surface, and reaches its maximal value for $N_T \rightarrow
\infty$ (practically, for $N_T=100$).

The eigenstates with small values of $l_H$ emerge in the energy
spectrum for {\it any} energy, thus resulting in a very slow
convergence to the limit of ergodicity. For example, the number of
eigenstates originated from $m=1$ is $N_{m=1}(E) =
\frac{2\pi}{d} \sqrt E$, therefore, the fraction of such
eigenstates is given by
$\frac{N_{m=1}(E)}{N(E)} = \frac{2b}{\pi^2} \frac{1}{\sqrt E} =
\sqrt{ \frac{4b}{\pi d} \frac{1}{N(E)}}$ where $N(E)$ is the total
number of states with energy less than $E$. One should
stress that in the energy spectrum these states appear regularily, due to
the expression Eq.(\ref{E00}) with $m=1$ and different $n$.

\begin{figure}
\hspace{-0.0cm}
\epsfxsize 8cm
\epsfbox{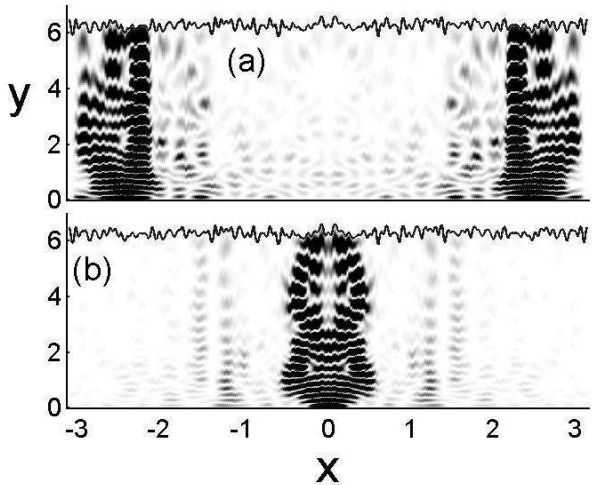}
\vspace{0.3cm}
\narrowtext
\caption{Two eigenstates that are strongly localized
in $x-$direction; (a) $\alpha=390$, (b) $\alpha=407$. The model
parameters are the same as in Fig. 1.}
\end{figure}

We would like to note that the type of localization
in the channel space we discuss here, is different from that studied 
for circular billiards with a rough surface
(see, for example, \cite{wigner}). The point is that
in our case classical diffusion in the
transverse momentum space turns out to be very strong compared with
quantum localazation effects \cite{long}. In contrast to circular billiards,
in our model with quasi-1D geometry the effects of a strong localization
(in the channel space) are due to the existence of a continuous set of classical
horisontal trajectories ("bouncing balls") which do not touch the rough boundary. 
As is shown in Ref. \cite{LFL98}, these trajectories result in anomalous
properties of conductance fluctuations for open waveguides of finite length.

It should be stressed that for any high energy one can find
eigenstates of a very different structure. To
demonstrate this fact, in Fig. 4 we report two typical examples
of strongly localized (in the coordinate $x$) eigenstates. 
These eigenstates are widely spanned in the 
unperturbed basis of $\hat H^0$, with large values $l_H\gg 1$, and 
they correspond to large values of $m \ll 1$.

To conclude with, we have analysed the structure of eigenstates 
of a quasi-1D waveguide with a rough surface, paying main attentsion
to their localization properties in the channel and coordinate
representaion. Different sets of strongly delocalized eigenstates
(along the waveguide) have been found, that have a quite specific 
form in the transverse direction. We have develop the approcah that
allows one to explain this form, using the transformation to 
new canonical variables.

Another result is that eigenstates turn out to have
very different localization properties and this
difference can not be treated as a result of fluctuations
only. Apart from the fluctuations, there are {\it regular} effects which are
due to a strong influence of the geometry. 
Namely, it was found that the eigenstates originated
from small values of $m$ are strongly localized (delocalized) 
in the channel (coordinate) representation,
and those associated with large $m$ are strongly delocalized (localized)
ones. This effect seems to be directly related 
to that found in Refs. \cite{SFYM99,IM03} for open finite waveguides,
where it was shown that characteristic scales for
scattering are different for different channels. Thus, it seems
questionable whether standard statistical approaches based on completely
random mathematical models, can adequately describe properties of
eigenstates.

This work was supported by the CONACyT (Mexico) Grant No. 34668-E.

\end{multicols}
\end{document}